\documentclass{appolb}
\usepackage{epsfig}
\begin{document}
\title{Neutrino propagation in dense hadronic matter
\thanks{Presented at XX Max Born Symposium
        ''Nuclear effects in neutrino interactions'',
        Wroclaw (Poland), December 7-10, 2005.}
       }%

\author{A. Rios and A. Polls
\address{Departament d'Estructura i Constituents de la Mat\`eria, \\
         Universitat de Barcelona, Avda. Diagonal 647, E-08028 Barcelona, Spain} 
\and
J. Margueron
\address{Institut de Physique Nucl\'eaire, Universit\'e Paris Sud F-91406 Orsay Cedex, France}
}

\maketitle

\begin{abstract}
Neutrino propagation in protoneutron stars requires the knowledge of the composition as well as the
dynamical response function of dense hadronic matter.
Matter at very high densities is probably composed of other particles than nucleons and little is known on
the Fermi liquid properties of hadronic multicomponent systems. We will discuss the effects that the presence of
$\Lambda$ hyperons might have on the response and, in particular, on its influence on the thermodynamical
stability of the system and the mean free path of neutrinos in dense matter.
\end{abstract}

\PACS{13.15.+g,13.75.Ev,21.30.Fe,25.30.Pt,26.60.+c}
  
Core collapse supernovae appear at the last stage of the evolution of massive stars \cite{bethe90}.
When the nuclear fuel of the star is exhausted, no more energy is produced in the interior and the
most internal iron shell collapses.
When the central density reaches the saturation density of nuclear matter $\rho_0$,
the collapsing matter bounces and forms a protoneutron star (PNS). This protostar is expected
to reach densities of the order of $2-3 \rho_0$ and temperatures up to $50$ MeV \cite{prakash97}.
The neutrinos produced during the collapse are trapped in the dense core. In addition,
the entropy increases and thus thermal production of neutrino pairs is enhanced. The PNS is then a
neutrino-rich environment, where neutrinos have been dynamically trapped.
However, neutrinos can diffuse out to the crust of the PNS, carrying the
heat of the inner core with them.
The early cooling process of the PNS is thus driven by neutrinos and it is very sensitive
to neutrino interactions in dense matter.
For instance, the details of the nucleon-nucleon (NN) interaction and the dynamical correlations in
the nuclear medium are of a capital importance for the cooling process \cite{margueron04}.
The presence of different
particles apart from nucleons (such as hyperons) can also influence the neutrino emissivity of the star,
either because of the effects that hyperons produce on the in-medium nucleons or
because of the neutrino interactions with hyperons. In any case, an important and illustrative
measure, which might be useful for early cooling process computations or supernovae evolution simulations
is the neutrino mean free path (MFP) in the hot and dense hadronic medium of the PNS.
The influence of the different baryons in this quantity has been previously studied, mainly
within relativistic mean field approaches \cite{reddy97}. 

From a theoretical point of view, Landau theory of Fermi liquids explains in a successful
way the properties of quantum liquids like $^3$He or nuclear matter \cite{baym78}.
Within Fermi Liquid Theory (FLT), the study of
the thermodynamical (TD) instabilities or the collective modes of hyper-nuclear systems could be
easily assessed. The main difficulty faced in the modeling of dense matter lies on our lack of knowledge
of the hyperon-nucleon (YN) and hyperon-hyperon (YY) interactions as well as on the in-medium modifications
of baryons.

The main idea underlying FLT simply states that, under certain assumptions,
the properties of Fermi liquids are only related to the excitations of the system
close to the Fermi surface. One can easily study these excitations (usually called
\emph{quasiparticles}) and find out the static and dynamic properties of the system.
In FLT, a fluctuation $\delta n_k = n_k - n_k^0$ of the Fermi surface will modify the
energy $E_0$ of the system in its ground state by an amount $\delta E = E[n_k] - E[n_k^0]$,
which is related with the fluctuation through:
\begin{eqnarray} 
\delta E &=& \sum_{\vec{k}_1} \epsilon(k_1) \delta n ( \vec{k}_1 )
+ \frac{1}{2} \sum_{\vec{k}_1,\vec{k}_2} f(\vec{k}_1,\vec{k}_2) \delta n(\vec{k}_1) \delta n(\vec{k}_2) \, ,
\end{eqnarray}
One can then define the quasiparticle energy, $\epsilon(k)$, as a first variation of the energy with
respect to the momentum distribution, while a second variation leads to identify
the function $f(\vec k_1, \vec k_2)$ to an interaction between quasiparticles:
\begin{eqnarray}
f(\vec{k}_1,\vec{k}_2) = \frac{\delta^2 E}{\delta n(\vec{k}_1) \delta n(\vec{k}_2) } \, .
\label{eq:lpars}
\end{eqnarray}
Since we are dealing with excitations around the Fermi surface, the
quasiparticle interaction is usually taken at $k_1=k_2=k_F$. The only remaining dependence is the one
in the Landau angle $\theta_{12}$ between $\vec k_1$ and $\vec k_2$, which can be accounted for by a
Legendre expansion:
\begin{eqnarray}
f(\theta_{12}) = \sum_{L} f_L P_L[ \cos(\theta_{12}) ] \, .
\end{eqnarray}
The Legendre coefficients $f_L$ are the so-called Landau parameters.
A relevant issue, already pointed out by Landau in his original derivation of FLT, is that these
parameters can be related to static and dynamical properties of the system, such as
the effective mass, the compressibility or the spin susceptibility.

Up to now we have restricted ourselves to a system of spinless fermions. When other quantum numbers,
such as spin are taken into account, the quasiparticle interaction should be properly generalized.
Usually, one expands the quasiparticle interaction in the following operational form:
\begin{eqnarray}
\mathcal{F}(\vec{k}_1 \sigma_1 ,\vec{k}_2 \sigma_2 ) 
&=& f(\vec{k}_1,\vec{k}_2)
+  g( \vec{k}_1,\vec{k}_2) \  \vec{\sigma}_1 \cdot \vec{\sigma}_2 \, .
\end{eqnarray}
Since $1$ and $\vec{\sigma}_1 \cdot \vec{\sigma}_2$ are the $S=0$ and $S=1$ projector operator in
the particle-hole channel, it is clear that $f$ and $g$ account for the $S=0$ and $S=1$ spin channels
of the particle-hole interaction.
Finally, whenever we deal with a system of different species of fermions (such as hadronic matter), we will
have to take into account the presence of the different Fermi surfaces and consider a Landau parameter
for each pair of species. In the case of a system of neutrons, protons and $\Lambda$'s, the possible $f$
parameters will be $ f^{nn}, f^{np}, f^{pp}, f^{n\Lambda}, f^{p\Lambda}, f^{\Lambda\Lambda} $.

As a first step in the application of FLT to hadronic systems, we will consider a simple case where
the in-medium hadronic interaction is of the Skyrme type. The NN, NY and YY in-medium interactions
will then be given by different sets of Skyrme-type parameterizations. 
Skyrme forces are very well known in the NN sector, where they have implied a major breakthrough in the
study of both the static properties and the collective motion of 
finite nuclei. In the fitting procedure of its parameters, one usually includes the properties of
nuclear matter and some given properties of medium and heavy spherical nuclei.
On the other hand, the available experimental data on heavy hyper-nuclei has
triggered the extension of this kind of forces to the YN and YY sector and, in particular, several 
parameterizations exist for the $\Lambda$N interaction \cite{lanskoy97}.
These Skyrme functionals have been usually fitted to the single-particle binding energies of $\Lambda$'s
in hyper-nuclei. To our knowledge, there is only one $\Lambda \Lambda$ Skyrme-type interaction,
fitted to the $\Lambda \Lambda$ binding in double hyper-nuclei \cite{lanskoy98}.
For convenience, from here on we will restrict ourselves to a single NN
parameterization, the SLy10 \cite{chabanat98} set of the Lyon group.
In the NY sector, we will use the LN1 parameterization of Ref.~\cite{lanskoy97},
while we will use the different YY parameterizations of Ref.~\cite{lanskoy98}.

\begin{figure}[t]
\begin{center}
\epsfig{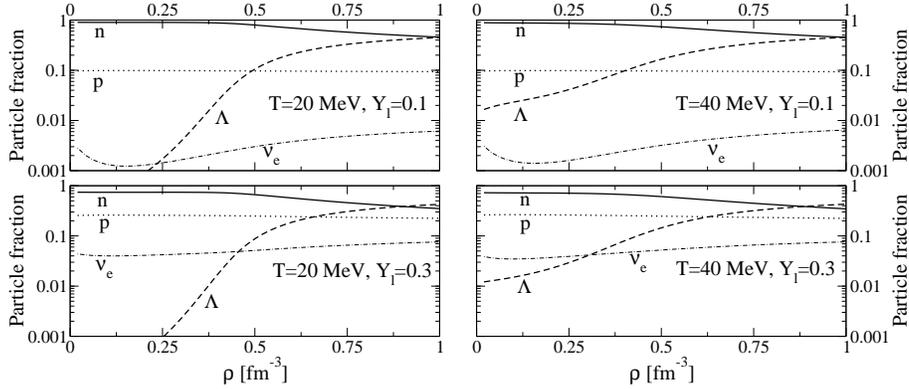}
\end{center}
\caption{$\beta$-stable composition of a PNS formed of neutrons, protons and $\Lambda$'s as a function of the
total baryonic density. The four panels correspond to two different temperatures, $T=20$ MeV (left panels) and
$T=40$ MeV (right panels), and two different leptonic fractions, $Y_l=0.1$ (upper panels) and $Y_l=0.3$ (lower panels).}
\end{figure}

We are interested in the description of hadronic matter in astrophysical environments. In a neutron 
star, the environment is defined by the $\beta$-equilibrium conditions, which arise from the assumption that
the weak reactions:
\begin{eqnarray}
&& p + e^- \leftrightarrow n + \nu_e  \\
&& p + e^- \leftrightarrow \Lambda + \nu_e \, 
\end{eqnarray}
are in equilibrium in the hadronic medium. 
In addition, a PNS is a charge neutral system ($\rho_p = \rho_e$).
In Fig.~1 we show some examples of the $\beta$-stable composition of protoneutron stars
at $T=20$ MeV (left panels) and $T=40$ MeV (right panels) temperatures as a function of density.
The upper (lower) panels correspond to a leptonic fraction of $Y_l=0.1$ ($Y_l=0.3$).
Different sets of NN and YN forces, however, would produce different particle fractions
and different astrophysical scenarios.

\begin{figure}[t]
\begin{center}
\epsfig{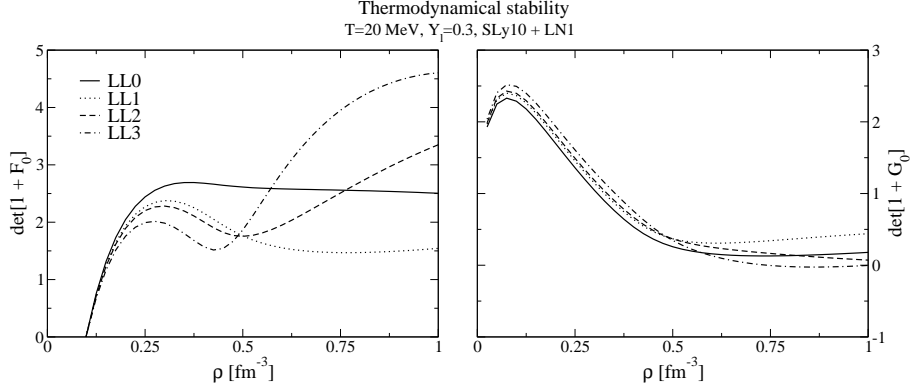}
\end{center}
\caption{Stability criteria for hadronic matter for the spin $S=0$ (right panel) and $S=1$ channels (left panel) as a
function of the total baryonic density. As illustrative examples, we have chosen a temperature and leptonic fraction
of $T=20$ MeV and $Y_l=0.3$, respectively.
Different sets of $\Lambda \Lambda$ interactions are shown with solid (LL0), dotted (LL1), dashed (LL2) and
dash-dotted lines (LL3).}
\end{figure}

As an example of the application of FLT to hadronic systems, we will compute the response function of
hadronic matter from the Landau parameters. If we consider diffusion reactions, in which neutrinos create 
a neutral particle-hole excitation on the system, the lowest-order response function is given in linear response 
theory by a $3 \times 3$ matrix structure:
\begin{eqnarray} 
\Pi_0=
\left(
\begin{array}{ccc}
\Pi_0^{nn} & 0 & 0 \\
0 & \Pi_0^{pp} & 0 \\
0 & 0 & \Pi_0^{\Lambda \Lambda} \\
\end{array} \right) \, ,
\end{eqnarray}
where each diagonal $\Pi_0^{ii}$ is given by the corresponding Lindhard function:
\begin{eqnarray}
\Pi_0^{ii}(\omega,\vec{q})&=& \int \, \frac{d^3 k}{(2\pi)^3} \frac{n_i[\epsilon(\vec{k})]-n_i[(\epsilon(\vec{k}+\vec{q})]}{\omega + \epsilon_i(\vec{k}) - \epsilon_i(\vec{k}+\vec{q}) + i \eta} \, .
\end{eqnarray}
The long range correlations are included through the Bethe-Salpeter equation:
\begin{eqnarray}
\Pi = \Pi_0 + \Pi_0 V_{res} \Pi  \, .
\end{eqnarray}
This is the well-known RPA approximation. The residual particle-hole interaction $V_{res}$ is
obtained from Eq.~(\ref{eq:lpars}) and reduced to the monopolar Landau parameters
$f$ ($g$) in the $S=0$ ($S=1$) channels. The Bethe-Salpeter equation
becomes then a coupled set of algebraic equations where the residual interaction is given by a $3 \times 3$ symmetric
matrix. In the $S=0$ channel, one finds:
\begin{eqnarray*}
\left(
\begin{array}{ccc}
f_0^{nn} & f_0^{np} & f_0^{n\Lambda} \\
f_0^{pn} & f_0^{pp} & f_0^{p\Lambda} \\
f_0^{\Lambda n} & f_0^{\Lambda p} & f_0^{\Lambda \Lambda} \\
\end{array} \right)  \, .
\end{eqnarray*}
The analytical expressions for the different $f$ and $g$ parameters can be found from the second derivatives of the 
energy Eq.~(\ref{eq:lpars}) and are explicitely given in \cite{mornas05}.
The solution of the Bethe-Salpeter equation is then:
\begin{eqnarray}
\Pi = \left( 1 -  \Pi_0 V_{res} \right)^{-1}  \Pi_0  \, .
\end{eqnarray}
If the determinant of the matrix $\left( 1 -  \Pi_0 V_{res} \right)$ becomes singular, the equation
might not be solvable. In the low energy limit, one can see that this determinant reduces to
the determinant of the TD stability matrix:
\begin{eqnarray*} 
\det \left[1 - V_{res} \Pi_0 \right]_{\omega \rightarrow 0} \rightarrow
\det \left(
\begin{array}{ccc}
\frac{\partial^2 E}{\partial \rho_n \rho_n} & \frac{\partial^2 E}{\partial \rho_n \rho_p}  & \frac{\partial^2 E}{\partial \rho_n \rho_{\Lambda}}  \\
\frac{\partial^2 E}{\partial \rho_p \rho_n} & \frac{\partial^2 E}{\partial \rho_p \rho_p}  & \frac{\partial^2 E}{\partial \rho_p \rho_{\Lambda}}   \\
\frac{\partial^2 E}{\partial \rho_{\Lambda} \rho_n} & \frac{\partial^2 E}{\partial \rho_{\Lambda} \rho_p}  & \frac{\partial^2 E}{\partial \rho_{\Lambda} \rho_{\Lambda}}  
\end{array} \right)  \, .
\end{eqnarray*}
A change of sign in this determinant signals the onset of a TD instability. Thus, the Landau parameters
of hadronic matter are related to its TD stability.

Results for the values of the low-energy limits of these determinants are shown in Fig.~2, where they are plotted
as a function of the total baryonic density in $\beta$-stable conditions. The different
sets of $\Lambda \Lambda$ interactions give different determinants due to both the changes in the $\beta$-stable
particle fractions and the differences on the parameters of the force. As a result, different TD
instabilities arise depending on the YY interaction. In particular, the onset of a ferromagnetic transition (signaled
by the zero of $\det [1+G_0]$) strongly depends on the set of forces that is used. In this case, all the forces
tend to a zero value at high densities (all of them are thus close to a ferromagnetic instability), but only
LL3 suffers from a real ferromagnetic transition at $\rho \sim 0.74$ fm$^{-3}$.

Once the Bethe-Salpeter equation is solved, we can find the structure functions
by means of the finite temperature relation (obtained from the detailed balance relationship):
\begin{eqnarray}
\mathcal{S}^{ij}_{(S)}(\omega,q) = - \frac{2}{1-\exp[-\omega/T]}\textrm{Im } \Pi^{ij}_{(S)}(\omega,q) \, ,
\end{eqnarray}
where $(S)$ denotes the two possible spin channels, $S=0$ and $S=1$.
In the calculation of the MFP we shall use the following ''generalized'' structure function,
where all the channels are taken into account by the following matrix product:
\begin{eqnarray}
\mathcal{S}_{(S)}=\left(c^n_{(S)} \  c^p_{(S)} \  c^{\Lambda}_{(S)} \right)
\left( \begin{array}{ccc}
\mathcal{S}^{nn}_{(S)} & \mathcal{S}^{np}_{(S)} & \mathcal{S}^{n\Lambda}_{(S)} \\
\mathcal{S}^{np}_{(S)} & \mathcal{S}^{pp}_{(S)} & \mathcal{S}^{p\Lambda}_{(S)} \\
\mathcal{S}^{n\Lambda}_{(S)} & \mathcal{S}^{p\Lambda}_{(S)} & \mathcal{S}^{\Lambda \Lambda}_{(S)}
\end{array} \right) 
\left( \begin{array}{c}
c^{n}_{(S)}  \\
c^{p}_{(S)}  \\
c^{\Lambda}_{(S)} 
\end{array} \right) 	\, ,
\label{eq:sf}
\end{eqnarray}
and where the $c^{i}_{(S)}$ are the corresponding weak charges of the particles which can be found in \cite{mornas05}.

\begin{figure}[t]
\begin{center}
\epsfig{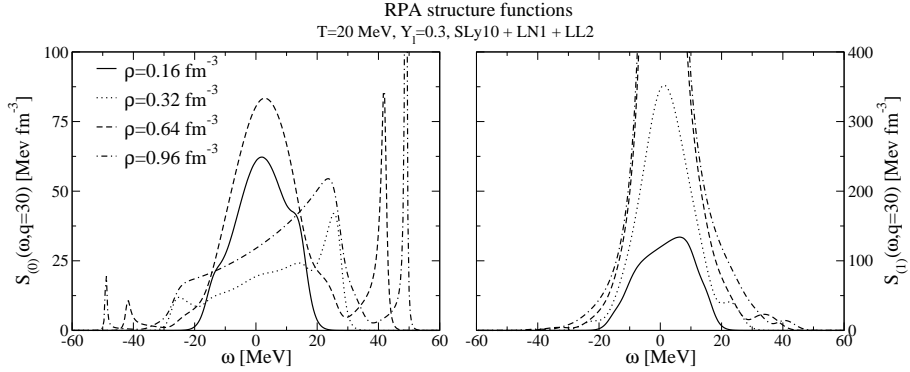}
\end{center}
\caption{Structure functions for the S=0 (right panel) and S=1 (left panel) channels as a function of energy for a
transferred momentum of $q=30$ MeV. The calculations have been performed in the same $\beta$-stable conditions as for
the previous figure at four different densities.}
\end{figure}

In the left (right) panels of Fig.~3 we show the structure function for the $S=0$ ($S=1$)
channel as a function of the energy, for a fixed momentum of $q=30$ MeV.
%The results have been computed for a $\beta$-stable composition at $T=20$ MeV and four different densities.
For the spin dependent channel, a collective mode around $\omega \sim 0$ is observed when the density
increases: this is again a signal of a close-by ferromagnetic transition.
One can see that when such a strong collective mode dominates the structure function, the neutrino MFP
is strongly suppressed. Thus, a different choice of NL and LL interactions which modifies the
TD instabilities of the system, might have a great influence on the neutrino MFP.

Finally, the mean free path is given by the following integral \cite{iwamoto82}:
\begin{eqnarray}
\frac{1}{\lambda} &=& 
\frac{G_F^2}{4 \pi^2} \int_{\infty}^{E_{\nu}} d\omega \frac{E'_{\nu}}{E_{\nu}} \int_{|\omega|}^{2E_{\nu}-\omega}
dq \, q \, [1-n(E'_{\nu})] \nonumber \\
&\times& \left[(1+\cos \theta) \mathcal{S}_{(0)}(\omega,q) + (3-\cos \theta) \mathcal{S}_{(1)}(\omega,q) \right] \, .
\label{eq:meanfp}
\end{eqnarray}
We show in Fig.~4 the results for the mean free path of neutrinos in dense matter for a leptonic
fraction of $Y_l=0.3$ and two temperatures, $T=20$ MeV and $T=40$ MeV. The left panels correspond to the
mean-field MFP computed from the non-correlated response functions $\Pi_0^{ii}$.
The solid lines correspond to $\beta$-stable matter composed by nucleons, electrons
and $\Lambda$'s, while the dashed lines correspond to matter without hyperons.
The neutrino energy is $E_\nu=3T$ MeV. For both temperatures, a clear reduction of the MFP is observed
when $\Lambda$'s are included in the system. This is due to the opening of a new channel of interaction,
with which neutrinos can interact, thus increasing their cross section (and reducing the
corresponding MFP).
The right panel of Fig.~4, on the other hand, illustrates the ratio between the MFP computed with RPA
correlations and the mean field MFP. The density ranges explored for each MFP correspond to the ranges
where no instabilities are observed for the corresponding $\beta$-stable matter.
A spin instability is observed for the matter formed by nucleons and electrons at densities above
$\rho \sim 0.5$ fm$^{-3}$. The presence of $\Lambda$'s pushes the ferromagnetic instability to higher
densities, and thus a non-zero neutrino MFP is observed in the high density range. This kind of effects,
where the presence of $\Lambda$'s in the system result in a suppression of the instabilities of matter,
are probably the most relevant ones for the neutrino propagation in hot and dense hadronic matter.

\begin{figure}[t]
\begin{center}
\epsfig{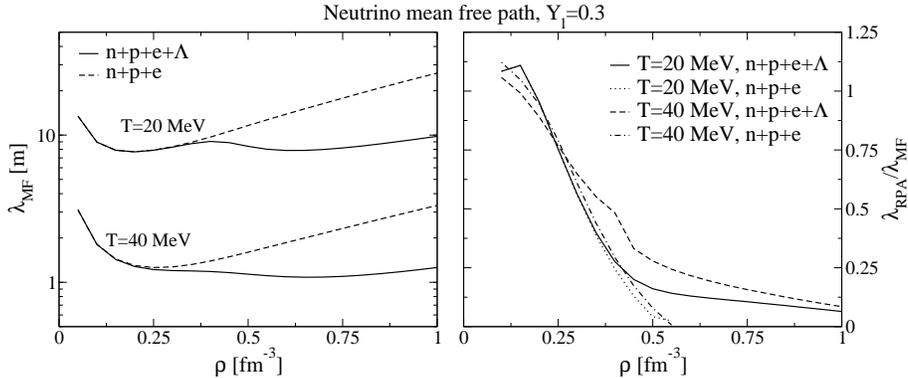}
\end{center}
\caption{Left panel: neutrino MFP in the mean field approximation for $\beta$-stable matter with $Y_l=0.3$.
The full (dashed) lines correspond to matter with (without) $\Lambda$'s.
%The upper lines have been computed at $T=20$ MeV, while the lower ones are at $T=40$ MeV.
Right panel: ratio of the RPA and mean field approximations of the neutrino MFP for the same conditions.}
\end{figure}

\end{document}